\documentclass[12pt]{article}
\begin{document}
\title{Ultra High Energy Behaviour}
\author{B.G. Sidharth\\
International Institute for Applicable Mathematics \& Information Sciences\\
Hyderabad (India) \& Udine (Italy)\\
B.M. Birla Science Centre, Adarsh Nagar, Hyderabad - 500 063
(India)}
\date{}
\maketitle
\begin{abstract}
We reexamine the behaviour of particles at Ultra High energies in
the context of the fact that the LHC has already touched an energy
of $7 TeV$ and is likely to attain $14 TeV$ by 2013/2014. A few
interesting possibilities are discussed.
\end{abstract}
\vspace{5 mm}
\begin{flushleft}
03.65.-W; 03.65.Pm
\end{flushleft}
\section{The "two component" Klein-Gordon Equation}
In relativistic Quantum Mechanics we encounter negative energies,
unlike in Classical Physics. In the case of the Dirac electron, this
lead to the postulation of the Hole theory. Let us now start with
the Klein-Gordon equation. In the relativistic Klein-Gordon
equation, the presence of the second time derivative and hence an
extra degree of freedom for the wave function $\psi$ lead to
interpretational problems, particularly in regard to negative energy
solutions. Pauli and Weiskopf overcame these difficulties by
treating the Klein-Gordon equation as describing a field and showed
that the two degrees of freedom of $\psi$ would be distinctly
charged states. Later Feshbach and Villars \cite{feshbach} showed
that this interpretation could also be carried over to the case of a
single particle description. As has been shown in detail by Feshbach
and Villars, we can rewrite the K-G equation in the Schrodinger
form, invoking a two component wave function,
\begin{equation}
\Psi = \left(\begin{array}{ll} \phi \\
\chi\end{array}\right),\label{2.16} \end{equation} The $K-G$
equation then can be written as (Cf.ref.\cite{feshbach} for details)
$$\imath \hbar (\partial \phi /\partial t) = (1/2m) (\hbar /\imath
\nabla - eA/c)^2 (\phi + \chi)$$
$$\quad \quad \quad +(e A_0 + mc^2)\phi,$$
\begin{equation}
\imath \hbar (\partial \chi / \partial t) = - (1/2m) (\hbar / \imath
\nabla - eA/c)^2 (\phi + \chi) + (e A_0 - mc^2)\chi\label{2.15}
\end{equation}
It will be seen that the components $\phi$ and $\chi$ are coupled in
(\ref{2.15}). In fact we can analyse this matter further,
considering free particle solutions for simplicity. We write, $$\Psi
= \left(\begin{array}{ll} \phi_0 (p) \\ \chi_0 (p)\end{array}\right)
\, e^{\imath / \hbar (p\cdot x-Et)}$$
\begin{equation}
\Psi = \Psi_0 (p)e^{\imath / \hbar (p\cdot x-Et)}\label{2.25}
\end{equation}
Introducing (\ref{2.25}) into (\ref{2.15}) we obtain, two possible
values for the energy $E$, viz.,
\begin{equation}
E = \pm E_p ; \quad E_p = [(cp)^2 +
(mc^2)^2]^{\frac{1}{2}}\label{2.26}
\end{equation}
The associated solutions are
$$\left.\begin{array}{ll} E = E_p  \quad \phi_0^{(+)} =
\frac{E_p+mc^2}{2(mc^2E_p)^{\frac{1}{2}}}\\
\psi_0^{(+)}(p): \quad \chi_0^{(+)} =
\frac{mc^2-E_p}{2(mc^2E_p)^{\frac{1}{2}}}\end{array}\right\}\phi_0^2
- \chi_0^2 = 1,$$
\begin{equation}
\left.\begin{array}{ll} E = - E_p  \quad \phi_0^{(-)} =
\frac{mc^2 - E_p}{2(mc^2E_p)^{\frac{1}{2}}}\\
\psi_0^{(-)}(p): \quad \chi_0^{(-)} = \frac{E_p +
mc^2}{2(mc^2E_p)^{\frac{1}{2}}}\end{array}\right\} \phi_0^2 -
\chi_0^2 = -1\label{2.27}
\end{equation}
It can be seen from this that even if we take the positive sign for
the energy in (\ref{2.26}), the $\phi$ and $\chi$ components get
interchanged with a sign change for the energy. Furthermore we can
easily show from this that in the non relativistic limit, the $\chi$
component is suppressed by order $(p / mc)^2$ compared to the $\phi$
component exactly as in the case of the Dirac equation \cite{bd}.
Let us investigate this circumstance further.\\
It can be seen that (\ref{2.15}) are Schrodinger equations and so
solvable. However they are coupled. We have from them,
\begin{equation}
\dot{\phi} + \dot{\chi} = (eA_0 + mc^2) (\phi + \chi) - 2mc^2
\chi\label{wD}
\end{equation}
In the case if
\begin{equation}
mc^2 > > eA_0 \quad (\mbox{or} \, A_0 = 0)\label{wF}
\end{equation}
(or in the absence of an external field) we can easily verify that
\begin{equation}
\phi = e^{\imath px-Et} \mbox{and} \, \chi = e^{\imath
px+Et}\label{wG}
\end{equation}
is a solution.\\
That is $\phi$ and $\chi$ belong to opposite values of $E (m \ne 0)$
(Cf. equation (\ref{2.27})). The above shows that the K-G equation
mixes the positive and negative energy solutions.\\
If on the other hand $m_0 \approx 0$, then (\ref{wD}) shows that
$\chi$ and $\phi$ are effectively uncoupled and are of same energy.
This shows that if $\phi$ and $\chi$ both have the same sign for
$E$, that is there is no mixing of positive and negative energy,
then the rest mass $m_0$ vanishes. A non vanishing rest mass
requires the mixing of both signs of energy. Indeed it is a well
known fact that for solutions which are localized, both signs of the
energy solutions are required to be superposed \cite{bd,schweber}.
This is because only positive energy solutions or only negative
energy solutions do not form a complete set. Interestingly the same
is true for localization about a time instant $t_0$.\\
That is physically, only the interval $(t_0 - \Delta t , t_0 +
\Delta t)$ is meaningful. This was noticed by Dirac himself when he
deduced his equation of the electron \cite{dirac}.\\
In any case both the positive and negative energy solutions are
required to form a complete set and to describe a point particle at
$x_0$ in the delta function sense. The narrowest width of a wave
packet containing both positive and negative energy solutions, which
describes the spacetime development of a particle in the familiar
non-relativistic sense, as is well known is described by the Compton
wavelength. As long as the energy domain is such that the Compton
wavelength is negligible then our usual classical type description
is valid. However as the energy approaches levels where the Compton
wavelength can no longer be neglected, then new effects involving
the negative energies and anti particles begin to appear (Cf.ref.\cite{feshbach}).\\
Further, we observe that from (\ref{wG})
\begin{equation}
t \to -t \Rightarrow E \to - E, \quad \phi \leftrightarrow
\chi\label{wJ}
\end{equation}
(Moreover in the charged case $e \to -e)$. This remark will be used
later using the fact that we are dealing with a two state system
(Cf.(\ref{wG})). On the other hand we will show that the Schrodinger
equation goes over to the Klein-Gordon equation if we allow $t$ to
move forward and also backward in $(t_0 - \Delta t, t_0 + \Delta
t)$. Here we have done the reverse of getting the Klein-Gordon
equation into two Schrodinger
equations. This is expressed by (\ref{2.15}).\\
In any case we would like to reiterate that the two degrees of
freedom associated with the second time derivative can be
interpreted, following Pauli and Weisskopf as positive and
negatively charged
particles or particles and anti particles.\\
We obtained from the Klein-Gordon equation a description in terms of
two Schrodinger equations. We will now show how (\ref{wJ}) with the
Schrodinger equation leads to the Klein-Gordon equation briefly
repeating
an earlier result.\\
We first define a complete set of base states by the subscript
$\imath \quad \mbox{and}\quad U (t_2,t_1)$ the time elapse operator
that denotes the passage of time between instants $t_1$ and $t_2$,
$t_2$ greater than $t_1$. We denote by, $C_\imath (t) \equiv <
\imath |\psi (t)
>$, the amplitude for the state $|\psi (t) >$ to be in the state $|
\imath >$ at time $t$. We have \cite{ijpap,cu,bgscsfqfst}
$$< \imath |U|j > \equiv U_{\imath j}, U_{\imath j}(t + \Delta t,t) \equiv
\delta_{\imath j} - \frac{\imath}{\hbar} H_{\imath j}(t)\Delta t.$$
We can now deduce from the super position of states principle that,
\begin{equation}
C_\imath (t + \Delta t) = \sum_{j} [\delta_{\imath j} -
\frac{\imath} {\hbar} H_{\imath j}(t)\Delta t]C_j (t)\label{2xe}
\end{equation}
and finally, in the limit,
\begin{equation}
\imath \hbar \frac{dC_\imath (t)}{dt} = \sum_{j} H_{\imath
j}(t)C_j(t)\label{2fe}
\end{equation}
where the matrix $H_{\imath j}(t)$ is identified with the
Hamiltonian operator. We have argued earlier at length that
(\ref{2fe}) leads to the Schrodinger equation
\cite{ijpap,bgscsfqfst}. In the above we have taken the usual
unidirectional time to deduce the non relativistic Schrodinger
equation. If however we consider a Weiner process in (\ref{2xe}) to
which we will return shortly, then we will have to consider instead
of (\ref{2fe})
\begin{equation}
C_\imath (t - \Delta t) - C_\imath (t + \Delta t) = \sum_{j}
\left[\delta_{\imath j} - \frac{\imath}{\hbar} H_{\imath
j}(t)\right] C_j (t)\label{2ge}
\end{equation}
Equation (\ref{2ge}) in the limit can be seen to lead to the
relativistic Klein-Gordon equation rather than the Schrodinger
equation with the second time derivative \cite{bgscsfqfst,tduniv}.
\section{Remarks and Discussion}
We summarize the following:\\
i) From the above analysis it is clear that a localized particle
requires both signs of energy. At relatively low energies, the
positive energy solutions predominate and we have the usual
classical type particle behaviour. On the other hand at very high
energies it is the negative energy solutions that predominate as for
the negatively charged counterpart or the anti particles. More
quantitatively, well outside the Compton wavelength the former
behaviour holds. But as
we approach the Compton wavelength we have to deal with the new effects.\\
ii) Let us remain in the realm of maximally localizable particles.
The point is that if we approach distances of the order of the
Compton wavelength, the negative energy solutions begin to dominate,
and we encounter the well known phenomenon of Zitterbewegung
\cite{dirac}. This modifies the coupling of the positive solutions
with an external field, particularly if the field varies rapidly
over the Compton wavelength. In fact this is the origin of the well
known Darwin term in the Dirac theory \cite{bd}. The Darwin term is
a correction to the interaction of the order
\begin{equation}
\left(\frac{p}{mc}\right)^4 \, \mbox{and} \,
\left(\frac{p}{mc}\right)^2\label{13}
\end{equation}
for spin $0$ and spin $1/2$ particles respectively.\\
iii) To reiterate if we consider the positive and negative energy
solutions given by $+ - E_p$, as in (\ref{2.27}), then we saw that
for low energies, the positive solution $\phi_0$ predominates, while
the negative solution $\chi_0$ is $\sim (\frac{v}{c})^2$ compared to
the positive solution. On the other hand at very high energies the
negative solutions begin to play a role and in fact the situation is
reversed with $\phi_0$ being suppressed in comparison to $\chi_0$.
This can
be seen from (\ref{2.27}).\\
iv) There is another elegant way in which we could look at the
considerations starting from (\ref{2.16}). In analogy with the iso
spin formulation, we could think of the wave function $\psi$ as
having two possible states in a charge-spin (or particle-anti
particle) space . In this case introducing the Pauli spin matrices
(Cf.\cite{feshbach}) given by
$$\tau_1 = \left(\begin{array}{ll} 0 \quad 1\\
1 \quad 0\end{array}\right); \quad \quad \imath \tau_2 =
\left(\begin{array}{ll} 0 \quad 1\\
-1 \quad 0\end{array}\right);$$
\begin{equation}
\tau_2 = \left(\begin{array}{ll} 1 \quad 0\\
0 \quad -1\end{array}\right); \quad \quad 1 =
\left(\begin{array}{ll} 1 \quad 0\\
0 \quad 1\end{array}\right)\label{2.17}
\end{equation}
the Hamiltonian (\ref{2.15}) can be written as
\begin{equation}
H = (\tau_3 + \imath \tau_2) (1/2m) (p-eA/c)^2 + mc^2 \tau_3 + e
\phi\label{2.18}
\end{equation}
Now we would have
\begin{equation}
\int \Psi^* \tau_3 \Psi d^3 x = \pm 1,\label{2.24a}
\end{equation}
Let us proceed further.\\
v) We have already seen the symmetry given in (\ref{wJ}): In case of
a charged particle, in addition, $e \to -e$ and vice versa (with
complexification). Furthermore it can be seen that the coordinate
$\vec{x}$, as it were splits into the coordinate $\vec{x}_1$ and
$\vec{x}_2$ which mimic the wave function in (\ref{2.16}) at low and
high energies, in the sense that the former dominates at low
energies while the latter dominates at high energies, following the
wave function as in (\ref{2.27}). The fact that these go into each
other following (\ref{wJ}) as $t \to -t$ can be explained in terms
of the development of a two Weiner process see briefly above
(Cf.\cite{tduniv}). In this case there are two derivatives, one for
the usual forward time and another for a backward time given by
\begin{equation}
\frac{d_+}{dt} x (t) = {\bf b_+} \, , \, \frac{d_-}{dt} x(t) = {\bf
b_-}\label{2ex1}
\end{equation}
where we are considering for the moment, a single dimension $x$.
This leads to the Fokker-Planck equations
$$
\partial \rho / \partial t + div (\rho {\bf b_+}) = V \Delta \rho
,$$
\begin{equation}
\partial \rho / \partial t + div (\rho {\bf b_-}) = - U \Delta
\rho\label{2ex2}
\end{equation}
defining
\begin{equation}
V = \frac{{\bf b_+ + b_-}}{2} \quad ; U = \frac{{\bf b_+ - b_-}}{2}
\label{2ex3}
\end{equation}
We get on addition and subtraction of the equations in (\ref{2ex2})
the equations
\begin{equation}
\partial \rho / \partial t + div (\rho V) = 0\label{2ex4}
\end{equation}
\begin{equation}
U = \nu \nabla ln\rho\label{2ex5}
\end{equation}
It must be mentioned that $V$ and $U$ are the statistical averages
of the respective velocities and their differences. We can then
introduce the definitions
\begin{equation}
V = 2 \nu \nabla S\label{2ex6}
\end{equation}
\begin{equation}
V - \imath U = -2 \imath \nu \nabla (l n \psi)\label{2ex7}
\end{equation}
We will not pursue this line of thought here but observe that the
complex velocity in (\ref{2ex7}) can be described in terms of a
positive or uni directional time $t$ only, but a complex coordinate
\begin{equation}
x \to x + \imath x'\label{2De9d}
\end{equation}
To see this let us rewrite (\ref{2ex3}) as
\begin{equation}
\frac{dX_r}{dt} = V, \quad \frac{dX_\imath}{dt} = U,\label{2De10d}
\end{equation}
where we have introduced a complex coordinate $X$ with real and
imaginary parts $X_r$ and $X_\imath$, while at the same time using
derivatives with respect
to time as in conventional theory.\\
We can now see from (\ref{2ex3}) and (\ref{2De10d}) that
\begin{equation}
W = \frac{d}{dt} (X_r - \imath X_\imath )\label{2De11d}
\end{equation}
That is we can either use forward and backward time derivatives and
the usual space coordinates as in (\ref{2ex3}) or we can use
derivatives with respect to the usual uni directional time
derivative
to introduce the complex coordinate (\ref{2De9d}) (Cf.ref.\cite{bgsfpl162003}.\\
Let us now generalize (\ref{2De9d}), which we have taken in one
dimension for simplicity, to three dimensions. Then we end up with
not three but four dimensions,
$$(1, \imath) \to (I, \tau),$$
where $I$ is the unit $2 \times 2$ matrix and $\tau$s are the Pauli
matrices. We get the special relativistic \index{Lorentz}Lorentz
invariant metric at the same time. (In this sense, as noted by Sachs
\cite{sachsgr}, Hamilton who made this generalization would have hit
upon \index{Special Relativity}Special Relativity, if he had
identified the new fourth coordinate
with time).\\
That is,\\
\begin{equation}
x + \imath y \to Ix_1 + \imath x_2 + jx_3 + kx_4,\label{A}
\end{equation}
where $(\imath ,j,k)$ now represent the \index{Pauli}Pauli matrices;
and, further,
\begin{equation}
x^2_1 + x^2_2 + x^2_3 - x^2_4\label{B}
\end{equation}
is invariant.\\
While the usual \index{Minkowski}Minkowski four vector transforms as
the basis of the four dimensional representation of the
\index{Poincare}Poincare group, the two dimensional representation
of the same group, given by the right hand side of (\ref{A}) in
terms of \index{Pauli}Pauli matrices, obeys the quaternionic algebra
of the second rank
\index{spin}spinors (Cf.Ref.\cite{bgsfpl162003,shirokov,sachsgr} for details).\\
To put it briefly, the \index{quarternion}quarternion number field
obeys the group property and this leads to a number system of
quadruplets as a minimum extension. In fact one representation of
the two dimensional form of the \index{quarternion}quarternion basis
elements is the set of \index{Pauli}Pauli matrices as in (\ref{A}).
Thus a \index{quarternion}quarternion may be expressed in the form
$$Q = -\imath \tau_\mu x^\mu = \tau_0x^4 - \imath \tau_1 x^1 - \imath \tau_2 x^2 -
\imath \tau_3 x^3 = (\tau_0 x^4 + \imath \vec \tau \cdot \vec r)$$
This can also be written as
$$Q = -\imath \left(\begin{array}{ll}
\imath x^4 + x^3 \quad x^1-\imath x^2\\
x^1 + \imath x^2 \quad \imath x^4 - x^3
\end{array}\right).$$
As can be seen from the above, there is a one to one correspondence
between a \index{Minkowski}Minkowski four-vector and $Q$. The
invariant is now given by
$Q\bar Q$, where $\bar Q$ is the complex conjugate of $Q$.\\
In this description we would have from (\ref{A})
\begin{equation}
[x^\imath \tau^\imath , x^j \tau^j] \propto \epsilon_{\imath jk}
\tau^k \ne 0\label{y}
\end{equation}
In other words, as (\ref{y}) shows, the coordinates no longer follow
a commutative geometry. It is quite remarkable that the
noncommutative geometry (\ref{y}) has been studied by the author in
some detail (Cf.\cite{tduniv}), though from the point of view of
Snyder's minimum fundamental length, which he introduced to overcome
divergence difficulties in Quantum Field Theory. Indeed we are
essentially in the same situation, because as we have seen, for our
positive energy description of the universe, there is the minimum
Compton wavelength cut off for a meaningful description
\cite{bgsextn,schweber,newtonwigner}.\\
Given (\ref{y}), it has been shown that the energy momentum relation
gets modified to
\begin{equation}
E^2 = p^2 + m^2 - \alpha l^2 p^4\label{Aa}
\end{equation}
The extra term in (\ref{Aa}) can be related to the Darwin term
(\ref{13}) (which shows moreover that $\alpha \sim 1$). In any case
for high energies or if $p > > 1$, then
$$ E^2 \sim - \alpha l^2 p^4 E$$
becomes imaginary! This is true if
$$\alpha l^2 p^4 > p^2 + m^2$$
that is if
$$p^2 > m^2 \, \, \mbox{so \, that}\, \, (\alpha \sim 1) p^2
\frac{1}{m^2} > 1 (l = \frac{1}{m}) \, \mbox{or} \, p^2 > m^2$$
which is true. All this happens when $O(l^2) \ne 0$ that is the
noncommutative geometry (\ref{y}) holds \cite{ijtp2004,uof}.\\
Let us write (\ref{2.15}) as (with $\hbar = 1 = c)$
\begin{equation}
H \phi = H_{11} \phi + H_{12} \chi\label{waee}
\end{equation}
and similarly we have
\begin{equation}
H \chi = H_{21} \phi + H_{22} \chi\label{wbee}
\end{equation}
We now observe that in Quantum Field Theory, a sub space of the full
Hilbert space can exhibit the complex or non Hermitian Hamiltonian.\\
Writing $H = M - \imath N$ where $M$ and $N$ are real \cite{wald} we
have
$$M_{11} = - \frac{\hbar^2}{2m} \nabla^2 + \frac{1}{2m}
\frac{e^2A^2}{c^2} + (e \phi + mc^2)$$
$$M_{21} = + \frac{\hbar^2}{2m} \nabla^2 - \frac{e^2A^2}{c^2} + (e
\phi - mc^2)$$
$$N_{11} = \frac{1}{m} \frac{eA}{c} \hbar \nabla = N_{12}$$
\begin{equation}
N_{21} = - N_{11} = N_{22}\label{wcee}
\end{equation}
We can now treat $|\phi , \chi >$ as a two state system and further
it follows from the above that
$$|\phi , \chi > (t) = exp (- N_{12} t) exp.$$
\begin{equation}
\quad \quad exp \left(- \imath M_{12} t\right) | \phi , \chi >
(0)\label{wdee}
\end{equation}
Equation (\ref{wdee}) shows that the states $|\phi >$ and $| \chi >$
decay, but decay at different rates (Cf. also \cite{wald}).\\
Treating $| \phi >$ and $| \chi >$ as particle and anti particle, we
have exactly this situation in $B$ and $K^0$ decay. It is remarkable
that this asymmetry has recently been observed in the case of $B$
mesons at the LHC by Sheldon Stone of Syracuse University and
co-workers. Such a behaviour was predicted by the author sometime
back \cite{bgsmod}. This has also been suggested in the case of
neutrinos and anti-neutrinos, in recent LHC observations. The point
here is that as in the case of the $B$ or $K^0$ mesons, the decay
rates of the particles and antiparticles would be different, thus
leading to a CPT violation. The above considerations provide an
explanation. This is also the case when equation (\ref{Aa}) holds:
the foregoing considerations
suggesting a similar explanation for the particle-anti particle asymmetry.\\
vii) We could now express the foregoing in the following terms: It
is well known that we get meaningful probability currents and
subluminal classical type situations using positive energy solutions
alone as long as we are at energies low enough such that we are well
outside the Compton scale. As we near the Compton scale however, we
begin to encounter negative energy solutions or these
anti-particles.\\
From this point of view, we can mathematically dub the solutions
according to the sign of energy $(p_0 |p_0|)$ of these states: $+1$
and $-1$. This operator commutes with all observables and yet is not
a multiple of unity as would be required by Schur's lemma, as it has
two distinct eigen values. This is a superselection principle or a
superspin with two states and can be denoted by the Pauli matrices.
The two states would refer to the positive energy solutions or our
ordinary universe and the negative energy solutions or what may be
called the anti-universe. The usual coordinates now take on a
quaternionic character and there is noncommutative spacetime.\\
The Compton wavelength is the Quantum Mechanical analogue of the
Einstein-Rosen bridge, in the sense that a penetration into this
region leads to opposite charges and what to our description would
be negative energy states and time going backwards. This "bridge"
connects our "positive energy" universe with what may be called an
anti universe that is one of negative energies. One of the puzzles
has been the asymmetry between matter and anti matter -- this could
be explained in the above terms of the decay caused at very high
energies.\\
In any case, ours is a universe that lies beyond the Compton scale
(or above the minimum extension) where the negative energy states
are irrelevant. We could think along the lines of $SU (2)$ and
consider the transformation \cite{taylor}
\begin{equation}
\psi (x) \to exp [\frac{1}{2} \imath g \tau \cdot \omega (x)] \psi
(x).\label{4.2}
\end{equation}
This leads to a covariant derivative
\begin{equation}
D_\lambda \equiv \partial_\lambda - \frac{1}{2} \imath g \tau \cdot
W_\lambda,\label{4.3a}
\end{equation}
as in the usual theory, remembering that $\omega$ in this theory is
infinitessimal. We are thus lead to vector Bosons $W_\lambda$ and an
interaction like the strong interaction, described by
\begin{equation}
W_\lambda \to W_\lambda + \partial_\lambda \omega - g \omega \Lambda
W_\lambda.\label{4.4}
\end{equation}
However we must bear in mind that this (non-electromagnetic)
interaction between particle and anti-particle \cite{report} would
be valid only within the Compton time, inside this Compton scale
Quantum
Mechanical bridge.\\
viii) Finally keeping in view the latest findings from LHC, it
appears that in the high energy p-p collisions, the product
particles display correlations. This has posed a puzzle. We could
describe this in what may be called a genetic model. We have to
think of the colliding particles as parents. They have genes or
information. These are spacetime related properties including for
example momenta, energy, relative locations, conservation laws and
so forth. The product particles are children who carry away some of
these properties in the form of correlations. In fact entanglement
could also come in the same category. However these "genetic bits of
information" cannot be equated with hidden variables because the
former are completely probabilistic.\\
Finally we would like to make the following remark: We have argued
that at ultra high energies, we have the Cini-Toushek transformation
rather than the low energy Foldy-Wothuysen transformation. The high
energy components in (\ref{2.16}) viz., $\chi$ dominate and the wave
function (\ref{2.16}) becomes as it were a two component neutrino
like wave function. There is thus handedness and a further
similarity is that the rest mass of the particle becomes very small
compared to the kinetic energy so that it approximates a massless
particle. So one could even speculate that some of the neutrino
sightings could well be heavier particles at luminal speeds.

\end{document}